\documentclass[doublecol]{epl2}
\usepackage{amsmath,amsfonts,bm}
\usepackage{graphicx}
\newcommand{\fneq}{\rho_t}
\newcommand{\feq}{\rho_t^{\rm eq}}
\newcommand{\frneq}{\hat \rho_t}
\newcommand{\freq}{\hat \rho_t^{\rm eq}}

\begin{document} 

\title{Dissipation and lag in irreversible processes}
\author{S Vaikuntanathan\inst{1,2} \and C Jarzynski\inst{1,3}}
\institute{\inst{1} Institute for Physical Science and Technology,University of Maryland, College Park, MD 20742\\ 
\inst{2} Chemical Physics Program, University of Maryland, College Park, MD 20742\\
\inst{3}Department of Chemistry, University of Maryland, College Park, MD 20742}

\abstract{
When a system is perturbed by the variation of external parameters,
a {\it lag} generally develops between the actual state of the system, $\rho_t$, and the equilibrium state corresponding to the current parameter values, $\rho_t^{\rm eq}$.
We establish a microscopic, quantitative relation between this lag and the {\it dissipated work} that accompanies the process.
We illustrate this relation using a model system.}

\maketitle 
\section{Introduction}

Irreversible thermodynamic processes are those that cannot be undone: the system of interest and its surroundings never return to their original states.
There are a number of attributes that we typically associate with such processes.
These include
(i) {\it dissipation} -- the dispersal of energy among many degrees of freedom;
(ii) {\it time-reversal asymmetry} -- the evident directionality of time's arrow; and
(iii) {\it broken equilibrium} -- either within the system of interest, or between it and its thermal surroundings.
For macroscopic systems these manifestations of irreversibility are related through the strict logic of the second law of thermodynamics.

For microscopic systems the second law must be interpreted statistically, making allowances for fluctuations around the mean behavior.
Far from being uninteresting, uninformative ``noise'', such fluctuations have in recent years been found to satisfy a number of exact and unexpected relations.~\cite{Bustamante2005}
These in turn have sharpened our understanding of the second law as it applies at the microscopic scale.
(See Section 7 of Ref.~\cite{CJ:Foundations} for a brief summary.)
Of specific relevance for the present paper is the discovery of quantitative relations between dissipation and time-reversal asymmetry, two of the above-mentioned manifestations of irreversibility.
While several such relations have appeared in the literature~\cite{Maes1999,Maes2003,Gaspard2004,CJ:Rareevents,KPV1}, we will focus on the formulation obtained by Kawai, Parrondo, and Van den Broeck~\cite{KPV1}, given by Eq.~\ref{original} below.

Consider a process in  which a system, initially at temperature $\beta^{-1}$, is driven away from equilibrium by varying an external parameter $\lambda$ from $A$ to $B$, over a time interval $0\le t\le\tau$.
By the second law, the average work performed on the system is no less than the free energy difference $\Delta F \equiv F_B - F_A$.
The excess $W_{\rm diss} \equiv \langle W \rangle - \Delta F \ge 0$
is eventually dissipated into the surroundings, and provides a physical measure of the dissipation that occurs during the process.
Now consider also the time-reversed process, in which the parameter is switched in the reverse manner from $B$ to $A$, and let $\rho_t \equiv \rho({\bf z},t)$ and $\tilde\rho_t \equiv \tilde\rho({\bf z},t)$ denote the phase space densities describing the evolution of the system during the two processes.
Since $\rho_t$ and $\tilde\rho_{\tau-t}$ represent statistical ``snapshots'' that correspond to the same value of the parameter $\lambda$~\cite{KPV1}, the relative entropy~\cite{Cover2006} between these two distributions, $D[\rho_t || \tilde\rho_{\tau-t}]$,
quantifies the extent to which the state of the system during the forward process ($A\rightarrow B$) is distinguishable from that during the reverse process ($B\rightarrow A$).
In other words, $D[\rho_t || \tilde\rho_{\tau-t}]$ provides an information-theoretic measure of time-reversal asymmetry.
By showing that
\begin{equation}
\label{original}
W_{\rm diss}  \ge \beta^{-1} \, D[\rho_t||\tilde \rho_{\tau-t}] ,
\end{equation}
Kawai {\it et al}~\cite{KPV1} have established a remarkably general inequality between these microscopic measures of (i) dissipation and (ii) time-reversal asymmetry.
The central goal of the present paper is to obtain a similarly general relation between dissipation and (iii) the loss of equilibrium during an irreversible process.

\begin{figure}[tbp]         
\includegraphics[scale=0.39,angle=0]{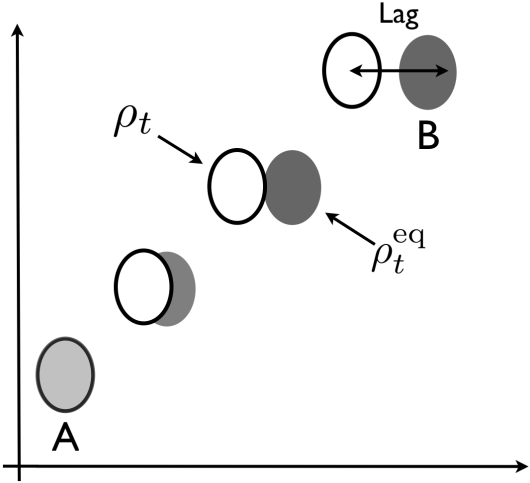} 
\caption{As the external parameter is switched from $A$ to $B$, a lag develops as the system pursues the equilibrium distribution corresponding to the changing external parameter.}  
 \label{lag.sketch}
\end{figure}

Restricting our attention to the forward process, let $\rho_t^{\rm eq}\equiv\rho^{\rm eq}({\bf z},\lambda_t)$ denote the equilibrium density corresponding to the value of the external parameter at time $t$.
Although the system begins in equilibrium ($\rho_0 = \rho_0^{\rm eq}$), at later times $\rho_t \ne \rho_t^{\rm eq}$.
This is illustrated schematically in Fig.~\ref{lag.sketch}: as $\lambda$ is varied with time, the system tries to keep pace with -- but ultimately lags behind -- the continually changing equilibrium distribution~\cite{Lag1}.
As in the previous paragraph, we can use relative entropy to quantify this lag:
$D[\rho_t ||\rho_t^{\rm eq}]$ measures the extent to which the system is out of equilibrium at time $t$.
The central result of this paper is the inequality
\begin{equation}
\label{alternate}
W_{\rm diss}(t)  \ge \beta^{-1}\, D[\rho_t ||\rho_t^{\rm eq}] ,
\end{equation}
where $W_{\rm diss}(t)$ is the amount of work dissipated up to time $t$ during the process.
Thus {\it the dissipated work dictates the maximum extent to which equilibrium can be broken} -- equivalently, the maximum amount of lag -- at a given instant during the process.

 {Our result Eq. \ref{alternate} is valid for $0\leq t\leq \tau$. However, unlike Eq.~\ref{original} our central result does not explicitly depend on the duration ($\tau$) of the process. }

We now derive our central result for systems driven arbitrarily far from equilibrium. We will then illustrate this result with an exactly solvable model system. 
\section{Theory}
\label{Theory}
 
Consider a classical system described by a parameter-dependent Hamiltonian $H_{\lambda}({\bf z})$ where ${\bf z}$ denotes a point in the phase space of the system.
At fixed parameter value $\lambda$ and temperature $\beta^{-1}$, the equilibrium state of the system is described by the probability distribution,
\begin{equation}
\label{equilibrium}
\rho^{\rm eq}({\bf z},\lambda) = \frac{e^{-\beta H_{\lambda}({\bf z})}}{Z_\lambda}
\end{equation}
with free energy $F_{\lambda}=-\beta^{-1}\ln Z_\lambda$.

Imagine a process during which the system is initially brought to thermal equilibrium with a heat bath at temperature $\beta^{-1}$, at fixed $\lambda=A$, after which the external parameter is varied from $\lambda(0)=A$ to $\lambda(\tau)=B$.
We will assume that the evolution of the system during this process is governed by dynamics that are Markovian and {\it balanced}; that is, the equilibrium distribution (Eq.~\ref{equilibrium}) is conserved when $\lambda$ is held fixed.
The time-dependent density $\rho_t = \rho({\bf z},t)$ describes an ensemble of trajectories evolving under these dynamics. 

For such processes, $\rho({\bf z},t)$ satisfies~\cite{CJ:MasterEquation, Hummer01,GEC2000} 
\begin{equation}
\label{phoequation}
\frac{e^{-\beta H_{\lambda(t)}({\bf z})}}{Z_A}= \rho({\bf z},t) \langle e^{-\beta W(t)} \rangle_{{\bf z},t} 
\end{equation}
where
\begin{equation}
\label{eq:workdef}
W(t) \equiv \int_0^t \dot{\lambda} \frac{\partial H_{\lambda}({\bf z}(t^\prime))}{\partial \lambda} dt^\prime
\end{equation}
denotes the work performed on the system along a trajectory ${\bf z}(t)$, and $\langle\dots\rangle_{{\bf z},t}$ denotes an average over all the trajectories that pass through ${\bf z}$ at $t$. Equation \ref{phoequation} can be rewritten using Eq \ref{equilibrium} to obtain
\begin{equation}
\label{phoequation1}
\frac{\rho({\bf z},t)}{\rho^{\rm eq}({\bf z},\lambda(t))} = \frac{e^{-\beta \Delta F(t)} }{\langle e^{-\beta W(t)} \rangle_{{\bf z},t}} ,
\end{equation}
where $\Delta F(t) = F_{\lambda(t)} - F_A$.
Taking the logarithm of both sides of this equation, then invoking Jensen's inequality~\cite{Cover2006}
\begin{equation}
\label{jensensineq}
\langle e^{-\beta W(t)} \rangle_{{\bf z},t} \ge e^{\langle -\beta W(t) \rangle_{{\bf z},t}} ,
\end{equation}
we get
\begin{equation}
\label{Jensen}
\langle W(t) \rangle_{{\bf z},t}-\Delta F(t) \ge \beta^{-1} 
\ln \frac{\rho({\bf z},t)}{\rho^{\rm eq}({\bf z},\lambda(t))}
\end{equation}
Finally, multiplying both sides of Eq \ref{Jensen} by $\fneq$ and integrating with respect to ${\bf z}$, we obtain
\begin{equation}
\label{result}
\langle  W(t) \rangle-\Delta F(t) \ge
\beta^{-1} \int d{\bf z} \, \rho_t \ln \frac{\rho_t}{\rho_t^{\rm eq}} .
\end{equation}
Since the left side of this equation represents the work dissipated to time $t$, and the right side is the relative entropy of $\rho_t$ with respect to $\rho_t^{\rm eq}$, we have arrived at our central result (Eq.~\ref{alternate}).

We now comment on a few aspects of this result.

First, Stein's lemma~\cite{Cover2006} relates the relative entropy $D[f||g] $ to the difficulty of statistically distinguishing between two distributions $f$ and $g$.
Thus, Eq \ref{result} directly connects the work dissipated up to a given time, $W_{\rm diss}(t)$, to a microscopic measure of the current deviation of the system from equilibrium, $D[\fneq||\feq]$.
It is worthwhile to discuss this deviation in some detail, for two separate situations.

(a) If the system remains in contact with a heat bath as $\lambda$ is switched from $A$ to $B$, then as suggested by Fig.~\ref{lag.sketch} we can picture the deviation of $\fneq$ from $\feq$ as a {\it lag} that develops because the system cannot keep pace with the changing equilibrium state.~\cite{Lag1,Lag2,Hermans91}
Eq.~\ref{result} tells us that the dissipated work places an upper bound on this lag.
In the special case that the parameter is varied quasistatically, and the heat bath is much larger than the system, then on general grounds we expect the system to remain in equilibrium, $\fneq=\feq$; in this case there is no dissipation, since $W(t) = \Delta F(t)$ for a reversible, isothermal process, and both sides of Eq.~\ref{result} are equal to zero.

(b) If we instead imagine that, after using a heat bath to prepare the system in an initial state of equilibrium, the heat bath is disconnected prior to the actual switching process, then during the interval $0\le t\le\tau$ the now-isolated system evolves under Hamilton's equations. 
As a result, a unique trajectory passes through any point ${\bf z}$ at time $t$, hence Eq.~\ref{jensensineq} becomes an equality and so does our central result:
\begin{equation}
\label{eq:equality}
W_{\rm diss}(t)  = \beta^{-1}\, D[\rho_t ||\rho_t^{\rm eq}] .
\end{equation}
Since the system is not continually attempting to equilibrate with an external heat bath, it is not immediately natural to view the deviation of $\fneq$ from $\feq$ in terms of lag.
(Indeed, even if $\lambda$ is varied quasistatically, the distribution $\fneq$ will deviate from the isothermal, canonical distribution $\feq$~\cite{CJ:MasterEquation,BiasedSampling,GECCJ07}.)
However, we can place this scenario within the ``lag framework'' by considering an isolated system to be a particular, limiting case of a system in contact with an external heat bath, in which the degree of thermal contact is so weak that the effects of the bath are negligible over a time interval of duration $\tau$.
If the external parameter is held fixed at $\lambda=B$ for $t>\tau$, then after a very long time the system does relax to a state of thermal equilibrium described by $\rho^{\rm eq}({\bf z},B)$.
In this paper we will adopt this perspective, and will view the relative entropy $D[\fneq||\feq]$ as a quantitative measure of lag, even in the case of a thermally isolated system.

We finally note that when $t=\tau$, Eqs.~\ref{original} and \ref{alternate} are equivalent, since $\tilde\rho({\bf z},0) = \rho^{\rm eq}({\bf z},B)$.
In other words, the initial state of the system during the reverse process is precisely the equilibrium state corresponding to the final value of $\lambda$ during the forward process.

\section{Examples}
\label{examples}

\begin{figure}[tbp]
\includegraphics[scale=0.4, angle=0]{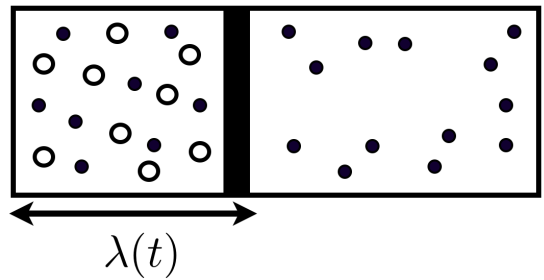}
\centering
\caption{A two-component dilute gas, where component 1 (open circles) is confined by the piston, while component 2 (filled circles) is not.}
\label{fig:rev:isothermal}
\end{figure}

Recent analyses of exactly solvable models~\cite{GPV1,dissipation_Jordan} have provided insight into Eq.~\ref{original}.
We now illustrate our central result, Eq.~\ref{alternate}, using a model that involves the quasistatic expansion or compression of a dilute
gas of particles in $d$ spatial dimensions.
The model, shown in Fig.~\ref{fig:rev:isothermal}, is motivated by Refs.~\cite{GECCJ07,CJ07:Website}. The gas is a two-component mixture, in which component 1 is confined by the piston (open circles in Fig.~\ref{fig:rev:isothermal}), while the particles of component 2 pass freely through the piston (filled circles).
Let $\lambda$ denote the position of the piston, $V_\lambda$ the volume of space to the left of the piston, $V$ the total volume of the container, and $N_1$ and $N_2$ the numbers of particles in each component.
For simplicity, we assume all particles have the same mass, $m$.

This mixture is initially allowed to come to thermal equilibrium with an external heat bath at temperature $\beta^{-1}$, with the piston held fixed at $\lambda=A$;
then thermal contact between the gas and the external bath is broken;
and finally, from $t=0$ to $t=\tau$, component 1 undergoes compression or expansion as the piston is manipulated quasistatically according to a protocol $\lambda(t)$.
During the latter stage the mixture evolves under Hamilton's equations in $2d(N_1+N_2)$-dimensional phase space.

This particular model is convenient because it can be used to illustrate both scenarios (a) and (b) discussed in the previous section.
If we define our system of interest to be the entire two-component mixture, then this model illustrates a system that is thermally isolated during the switching process, as per scenario (b).
Alternatively, if we take our system of interest to be component 1, and view component 2 as part of a heat bath, then the model illustrates scenario (a).
We will analyze these two cases below.
We will solve explicitly for dissipated work and relative entropy in each case, and will show that our central result is an equality in the case of a thermally isolated system of interest (Eq.~\ref{eq:equality}), and an inequality when the system remains in contact with a heat bath (Eq.~\ref{alternate}).

\subsection{Hamiltonian Dynamics}
\label{subsec:hamiltonian}

Let  ${\bf z} \equiv \{{\bf z}_1,{\bf z}_2\}$ denote a point in the full, $2 d (N_1+N_2)$-dimensional phase space, with ${\bf z}_1$ and ${\bf z}_2$ denoting the phase coordinates of components 1 and 2, respectively.
The Hamiltonian for this system is $H_\lambda({\bf z})$.
As in Ref.~\cite{GECCJ07}, we take the term ``dilute gas'' to imply that, while particles do exchange energy via pairwise collisions, the mean free path between collisions is much greater than the characteristic distance between nearby particles.
For practical purposes, we take this to mean that the particle-particle interaction terms in $H_\lambda({\bf z})$ can be neglected in the calculations that follow.
Thus $H_\lambda$ is taken to be a sum of kinetic energies, and hard-wall potentials that confine the two components to volumes $V_\lambda$ and $V$.
We also assume that when the piston is held fixed, the Hamiltonian dynamics are ergodic, i.e.\ the mixture is able to self-equilibrate via particle-particle collisions.
Finally, the term ``quasistatic'' is meant to imply that the compression or expansion proceeds sufficiently slowly for continual self-equilibration to occur.

For fixed $\lambda$ and positive energy value $E$, let $\phi_\lambda(E)$ denote the volume of phase space enclosed by the energy shell (i.e.\ surface of constant energy) $H_\lambda({\bf z}) = E$; and let us think of $g_\lambda(E) \equiv \partial\phi/\partial E$ as the ``surface area'' of this shell.
By explicit calculation, we have
\begin{eqnarray}
\label{phi}
\phi_\lambda(E) &=& \int d{\bf z} \, \theta (E - H_\lambda)=
\mu^k V_\lambda^{N_1} V^{N_2}
\frac{E^k}{k\Gamma(k)} \\
\label{gev}
g_\lambda(E) &=& \int d{\bf z} \, \delta( E - H_\lambda) =
\mu^k V_\lambda^{N_1} V^{N_2}
\frac{E^{k-1}}{\Gamma(k)}
\end{eqnarray}
where
\begin{equation}
k = \frac{d}{2}(N_1+N_2)
\quad,\quad
\mu = 2\pi m \quad ,
\end{equation}
and $\Gamma(k)$ is the gamma function.
At temperature $\beta^{-1}$, the partition function and free energy of the mixture are:
\begin{subequations}
\label{eq:ZF}
\begin{align}
\label{eq:Z}
Z_\lambda(\beta) &= \int dE\,g_\lambda \, e^{-\beta E} =
\mu^k V_\lambda^{N_1} V^{N_2} \beta^{-k} \\
F_\lambda(\beta) &= -\beta^{-1} \ln Z_\lambda  \quad .
\end{align}
\end{subequations}

When the piston is  moved quasistatically from $\lambda(0)=A$ to $\lambda(\tau)=B$, the value of $\phi_\lambda(H_\lambda)$ is an adiabatic invariant~\cite{GECCJ07}.
By Eq.~\ref{phi}, this implies
\begin{equation}
\label{invariance}
V_A^{N_1} E_0^k = V_{\lambda(t)}^{N_1} E_t^k
\end{equation} 
along a trajectory ${\bf z}(t)$ with energy $E_t\equiv H_{\lambda(t)}({\bf z}(t))$.
The work performed on the mixture is given by net change in its energy,
\begin{equation}
\label{changeenergy}
W(t) = E_t-E_0 =\left[ \frac{V_A^{N_1/k}}{V_{\lambda(t)}^{N_1/k}}-1\right] E_0 \equiv \alpha(t) E_0 .
\end{equation}
Since $W(t)$ is determined uniquely by the initial energy, $E_0$, and initial conditions are sampled from the equilibrium distribution at temperature $\beta^{-1}$, we have:
\begin{equation}
\label{averageW}
\begin{split}
\langle W(t) \rangle & = \frac{1}{Z_A} \int_0^\infty dE_0  g_A(E_0) e^{-\beta E_0} \alpha(t) E_0   \\
&=
\alpha(t) \, \langle E_0\rangle = k\beta^{-1} \alpha(t) \quad .
\end{split}
\end{equation}
Finally, from Eq.~\ref{eq:ZF} we get
\begin{equation}
\label{eq:deltaF}
\Delta F(t) = N_1\beta^{-1}\ln\frac{V_A}{V_{\lambda(t)}}
= k\beta^{-1} \ln \left[ \alpha(t) + 1 \right] \quad.
\end{equation}
From the first expression on the right is is clear that this quantity depends on $N_1$ but not on $N_2$; effectively, $\Delta F(t)$ specifies a free energy difference between two equilibrium states of component 1, as the equilibrium state of component 2 is unaffected by the piston.

Combining Eqs.~\ref{averageW} and \ref{eq:deltaF} yields the following compact expression for the dissipated work:
\begin{equation}
\label{eq:wtdiss}
W_{\rm diss}(t) =
k \beta^{-1}
\left[
\alpha - \ln(\alpha+1)
\right] \quad.
\end{equation}

To compute $D[\fneq||\feq]$, we consider a trajectory ${\bf z}_t \equiv {\bf z}(t)$ evolving under Hamilton's equations.
By Liouville's theorem, the value of phase space density is conserved along this trajectory, hence
\begin{equation}
\label{eq:liouville}
\rho({\bf z}_t,t) = \rho({\bf z}_0,0)
=
\frac{1}{Z_A(\beta)} e^{-\beta E_0}
=
\frac{1}{Z_A(\beta)} e^{-\bar\beta_t E_t}
\quad,
\end{equation}
where $\bar\beta_t = \beta / [\alpha(t)+1]$, and we have made use of Eq.~\ref{changeenergy}.
With Eq.~\ref{eq:Z} we can confirm that $Z_A(\beta) = Z_{\lambda(t)}(\bar\beta_t)$, thus
\begin{equation}
\label{eq:rho_t}
\rho({\bf z},t) = \frac{1}{Z_{\lambda(t)}(\bar\beta_t)} e^{-\bar\beta_t H_{\lambda(t)}({\bf z})} \quad.
\end{equation}
In other words, during the compression or expansion process the phase space density is a canonical distribution with a slowly time-dependent temperature, $\bar\beta_t^{-1}$.
By contrast, $\rho^{\rm eq}$ is defined at a constant temperature,
\begin{equation}
\label{eq:rho_eq}
\rho^{\rm eq}({\bf z},\lambda(t)) = \frac{1}{Z_{\lambda(t)}(\beta)} e^{-\beta H_{\lambda(t)}({\bf z})}.
\end{equation}
We therefore have
\begin{equation}
\ln \frac{\rho_t}{\rho_t^{\rm eq}} =
\left( \beta - \bar\beta_t \right) H_{\lambda(t)}({\bf z}) - k \ln \left( \beta / \bar\beta_t \right) .
\end{equation}
Multiplying both sides by Eq.~\ref{eq:rho_t} and integrating, we get
\begin{equation}
\begin{split}
D[\fneq||\feq]
&= \left( \beta - \bar\beta_t \right) k \bar\beta_t^{-1} - k \ln (\alpha+1) \\
&= k \left[
\alpha - \ln(\alpha+1)
\right] \quad.
\end{split}
\end{equation}
Comparing with Eq.~\ref{eq:wtdiss}, we see that Eq.~\ref{eq:equality} is satisfied.

\subsection{Stochastic dynamics}
\label{subsec:stochastic}

Now let us view component 1 of our mixture as the system of interest, and component 2 as part of the heat bath.~\footnote{
Thus the entire heat bath is composed of both the external bath used to prepare the initial state of equilibrium, and the particles of component 2, which remain in contact with the system of interest during the process.}
The phase space of the system of interest is now $2dN_1$-dimensional, and evolution in this space is stochastic rather than deterministic, as the variables ${\bf z}_2$ have been projected out.
We will use a carat ($\,\hat\,\,$) to denote reduced phase space densities describing the system of interest (component 1):
\begin{equation}
\begin{split}
\hat\rho_t &= \hat\rho({\bf z}_1,t) = \int d{\bf z}_2 \, \rho({\bf z},t) \\
\hat\rho_t^{\rm eq} &= \hat\rho^{\rm eq}({\bf z}_1,\lambda(t)) = \int d{\bf z}_2 \, \rho^{\rm eq}({\bf z},\lambda(t))
\quad.
\end{split}
\end{equation}
The relative entropy
$D\left[ \hat\rho_t || \hat\rho_t^{\rm eq} \right]$
quantifies the degree to which the system of interest is out of equilibrium
(as before, ``equilibrium'' is defined by the temperature $\beta^{-1}$ and the current value of $\lambda$)
and we wish to compare this with the dissipated work, $W_{\rm diss}(t)=\langle W(t)\rangle - \Delta F(t)$.

(Before proceeding further, we note that the stochastic evolution of the system of interest is non-Markovian, thus it is not immediately obvious that the analysis of the previous section (Theory) can be applied to this situation; see the assumptions stated after Eq.~\ref{equilibrium}.
To address these concerns, we verify in the Appendix that Eq.~\ref{phoequation1} remains valid for the reduced densities, even though the evolution is non-Markovian.)

Since the particles of component 2 pass freely through the piston, the values of $\langle W(t)\rangle$ and $\Delta F(t)$ are the same as before (see comment following Eq.~\ref{eq:deltaF}).
By contrast, since the reduced densities are obtained by projecting from the full phase space to that of component 1, there will be a reduction in the value of the relative entropy~\cite{Cover2006}: $D\left[ \hat\rho_t || \hat\rho_t^{\rm eq} \right] < D\left[ \rho_t || \rho_t^{\rm eq} \right]$, as we now confirm by direct evaluation.

Because $\rho_t$ and $\rho_t^{\rm eq}$ are canonical distributions in the full phase space (Eqs.~\ref{eq:rho_t}, \ref{eq:rho_eq}), the reduced densities are also canonical:
\begin{eqnarray}
\label{eq:rhohat_t}
\hat\rho_t &=& \frac{1}{\hat Z_{\lambda(t)}(\bar\beta_t)}
e^{-\bar\beta_t H_{\lambda(t)}^{(1)}({\bf z}_1)} \\
\hat\rho_t^{\rm eq} &=& \frac{1}{\hat Z_{\lambda(t)}(\beta)}
e^{-\beta H_{\lambda(t)}^{(1)}({\bf z}_1)} ,
\end{eqnarray}
where $H^{(1)}$ is the Hamiltonian for component 1, and
\begin{equation}
\hat Z_\lambda(\beta) = \mu^{k_1} V_\lambda^{N_1} \beta^{-k_1}
\quad,\quad
k_1= dN_1/2 .
\end{equation}
We now have
\begin{equation}
\ln \frac{\hat\rho_t}{\hat\rho_t^{\rm eq}} =
\left( \beta - \bar\beta_t \right) H_{\lambda(t)}^{(1)}({\bf z}_1) - k_1 \ln \left( \beta / \bar\beta_t \right) .
\end{equation}
Multiplying by Eq.~\ref{eq:rhohat_t} and integrating, we obtain
\begin{equation}
\label{eq:reduced_relative_entropy}
\begin{split}
D[\hat\rho_t || \hat\rho_t^{\rm eq}]
&= \left( \beta - \bar\beta_t \right) k_1 \bar\beta_t^{-1} - k_1 \ln (\alpha+1) \\
&= k_1 \left[
\alpha - \ln(\alpha+1)
\right] \\
&= \frac{N_1}{N} \beta W_{\rm diss}(t) = \frac{N_1}{N} D[\fneq||\feq] ,
\end{split}
\end{equation}
where $N = N_1+N_2$ is the total number of particles in the mixture.~\footnote{
Eq.~\ref{eq:reduced_relative_entropy} is easy to understand:
$D[\fneq||\feq]$ is a sum of equal contributions from each of the $N$ particles in the mixture, but only $N_1$ particles contribute to $D[\frneq||\freq]$.
}
As expected, our central result (Eq.~\ref{alternate}) now holds as a strict inequality.

\begin{figure}[tbp]
\includegraphics[scale=0.7, angle=0]{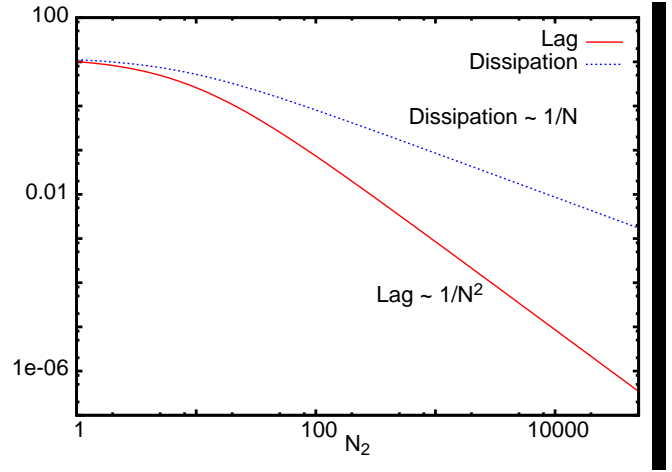}
\centering
\caption{Dissipation ($\beta W_{\rm diss}(t)$) and lag ($D[\frneq||\freq]$) are plotted as functions of $N_2$, with $N_1=10$, $d=3$, $V_0/V_{\lambda(t)}=5$, and $\beta=1$. The isothermal limit is achieved as $N_2 \to \infty$.}
\label{fig:rev:dissipationvlag}
\end{figure}

Finally, let us consider what happens when component 2 is much larger than component 1; formally, $N_2 \rightarrow\infty$ with $N_1$ fixed.
By straightforward evaluation we find
\begin{equation}
\begin{split}
\bar\beta_t &= \beta + {\it O}(1/N) \\
W_{\rm diss}(t) &\sim 1/N \\
D[\hat\rho_t || \hat\rho_t^{\rm eq}] &\sim 1/N^2.
\end{split}
\end{equation}
Physically, this limit describes the reversible {\it and isothermal} compression or expansion of component 1, with component 2 playing the role of an infinite heat bath.
We see that both $W_{\rm diss}(t)$ and $D[\hat\rho_t || \hat\rho_t^{\rm eq}]$ approach zero, but at different rates, as illustrated in Fig.~\ref{fig:rev:dissipationvlag}.

\section{Summary}
\label{conclusion}

When a system is driven away from equilibrium by the variation of external parameters, the relative entropy $D[\fneq||\feq]$ quantifies the degree to which the current state of the system, $\rho({\bf z},t)$, lags behind the instantaneous equilibrium state, $\rho^{\rm eq}({\bf z},\lambda(t))$.
Our central result, Eq.~\ref{result}, shows that the dissipated work, $W_{\rm diss}(t)$, provides an upper bound on the value of this lag.
In the special case that the dynamics of the system are Hamiltonian, the dissipation fully specifies the lag (Eq~\ref{eq:equality}).
These results complement analogous results obtained for the relationship between dissipated work and time-reversal asymmetry~\cite{KPV1}, as measured by $D[\rho_t || \tilde\rho_{\tau-t}]$.

\acknowledgements
We gratefully acknowledge useful discussions with Jordan Horowitz and Andy Ballard, and financial support from the National Science Foundation under CHE-0841557 and the University of Maryland, College Park. 

\vskip .2in

\section{Appendix}
\label{appendix}

In the full phase space, Eq.~\ref{phoequation} can be rewritten as
\begin{equation}
\rho^{\rm eq}({\bf z},\lambda(t)) e^{-\beta \Delta F(t)} = \rho({\bf z},t) e^{-\beta W(t)} ,
\end{equation}
where $W(t)$ is the work performed along the unique trajectory that passes through ${\bf z}$ at time $t$.
Integrating both sides with respect to ${\bf z}_2$,  we get
\begin{equation}
\hat\rho^{\rm eq}({\bf z}_1,\lambda(t)) e^{-\beta \Delta F(t)} = \hat\rho({\bf z}_1,t) \left\langle e^{-\beta W(t)}\right\rangle_{{\bf z}_1,t} .
\end{equation}
Rearranging terms we see that the reduced densities satisfy Eq.~\ref{phoequation}.

\end{document}